\documentstyle[pre,preprint,aps]{revtex}
\tightenlines
\newcommand{\siml}{\stackrel{<}{\sim}}

\begin{document}
\draft

\title{
Spike-train responses of a pair of Hodgkin-Huxley neurons  
coupled by excitatory and inhibitory synapses \\
and axons with time delay 
}
\author{
Hideo Hasegawa\footnote{E-mail: hasegawa@u-gakugei.ac.jp}
}
\address{
Department of Physics, Tokyo Gakugei University,
Koganei, Tokyo 184-8501, Japan
}
\date{\today}
\maketitle
\begin{abstract}

\end{abstract}
Numerical calculations have been made on the spike-train response of 
a pair of Hodgkin-Huxley (HH) neurons coupled by synapses and
axons with time delay. 
The recurrent excitatory-excitatory,
inhibitory-inhibitory,  excitatory-inhibitory,
and inhibitory-excitatory couplings are adopted.
The coupled, excitable HH neurons are assumed to receive
the two kinds of spike-train inputs: 
the transient input consisting of
$M$ impulses for the finite duration ($M$: integer)
and the sequential
input with the constant interspike interval (ISI).
The HH neurons in all the kinds of couplings are found to
play a role of memory storage with on-off switching.
When the coupling strength and the time delay
are changed, the distribution of the 
output ISI $T_{\rm o}$ shows bifurcation (multifurcation), 
metastability and chaotic behavior. The calculation
of the time correlation shows that the synchronization 
between the two HH neurons
is  well preserved even when the distribution
of their $T_{\rm o}$ is chaotic.
The correlation dimension of the
cycles of $T_{\rm o}$ is shown to depend not only
on the model parameters but also on the type of input ISIs.

\noindent
\vspace{0.5cm}
\pacs{PACS No. 87.18.Sn 84.35.+i }
%

\begin{center}
{\bf I. INTRODUCTION}
\end{center}

Neurons communicate by producing sequence of action 
potentials or spikes. It has been widely believed 
that information is encoded in the average rate of firings, 
the number of action potentials over some suitable 
intervals.
This firing rate hypothesis was first proposed by 
Andrian\cite{Andrian26} from a study of frog, in which
the firing rate monotonically increases with an increase
of the stimulus strength.
By applying the firing rate hypothesis, the properties
of many types of neurons in brain have been investigated
and the theoretical models have been 
developed \cite{Hopfield82}.

When all action potentials are taken to be identical
and only the times of firing of a given neuron
are considered, we obtain a discrete series of times,
$\{t_{n}\}$, which is expected to contain the information.
In the {\it rate coding}, only the average of the rate of
the interspike interval (ISI) is taken into account, and
then some or most of this information is neglected.

In recent years, the alternative {\it temporal coding},
in which detailed spike timing is taken to play an important role,
is supported by experiments in a variety of 
biological systems: 
sonar processing of bats \cite{Suga83}, 
sound localization of owls \cite{Konishi92},
electrosensation in electric fish \cite{Carr86}, 
visual processing
of cats \cite{Eckhorn88} \cite{Gray89}, 
monkeys \cite{Rolls94} and human \cite{Thorpe96}.
It is now primarily important to understand what
kind of code is employed in biological systems:
rate code, temporal code or others\cite{Rieke96}\cite{Maass98}.

Neural functions are performed in the activity of neurons.
Since the Hodgkin-Huxley (HH) model was proposed to account for
the squid giant axon \cite{Hodgkin52}, 
its property has been intensively
investigated. 
Its responses to an applied dc 
\cite{Nemoto76}-\cite{Guttman80}
and sinusoidal currents\cite{Aihara84}\cite{Matsumoto84} 
have been studied.
The HH-type models have been widely employed for a study on 
activities of {\it transducer neurons} such as motor
and relay neurons, 
which transform
amplitude-modulated inputs to spike-train outputs.
Regarding the single HH neuron as a {\it data-processing neuron},
the present author \cite{Hasegawa00} 
(referred to as I hereafter) recently investigates its response
to the spike-train inputs whose ISIs
are modulated by deterministic, semi-deterministic (chaotic)
and stochastic signals.

Several investigations have been reported on
the property of a pair of the HH neurons
\cite{Vreeswijk94}-\cite{Kanamaru98}.
In the network of two HH oscillators coupled 
by excitatory couplings without time delay, 
the unit fires periodically in the
synchronized state.
It is, however, not the case when the excitatory
couplings have some time delay,
for which the antiphase state becomes more stable
than the synchronized state \cite{Vreeswijk94}.
Rather, inhibitory couplings with substantial time delay
lead to the in-phase synchronized states
in the coupled HH oscillators \cite{Vreeswijk94}.
The similar conclusion is obtained also in the coupled
integrate-and-fire (IF) oscillators \cite{Vreeswijk94}
\cite{Ernst98}
\cite{Ernst95}-\cite{Coobes97}.
The phase diagrams for the synchronized state
and various cluster states in the coupled HH oscillators
are obtained as functions of the synapse
strength and the time delay 
\cite{Park96}-\cite{Kim98}.

In recent years, much attention has been paid
to the delayed-feedback systems described by the 
delay-differential equation (DDE)\cite{Ikeda82}-\cite{Fisher94}.
Their property has been investigated with the use of
various functional forms for the delay-feedback term in DDE. 
The exposed properties include 
the odd-harmonic solutions \cite{Ikeda82}\cite{Ikeda87},
the bifurcation (multifurcation) leading to chaos
\cite{Ikeda87}-\cite{Mork90}, 
the multistability \cite{Ikeda87},
and the chaotic itinerancy \cite{Otsuka90}-\cite{Fisher94}.
Among them the multistability is intrigue because it
may be one of conceivable mechanisms for memory storage
in biological neural networks.
It has been shown by Ikeda and Matsumoto\cite{Ikeda87} 
that when the delay time is larger
than the response time in the delayed feedback system,
information may be stored in temporal patterns. 
Actually, Foss, Longtin, Mensour and Milton \cite{Foss96}
demonstrate this ability in the coupled HH (and IF) neurons
with the time-delayed feedback.

Recurrent loops involving two or more neurons with 
excitatory and/or inhibitory synapses 
are found in biological systems
such as hippocampus \cite{Freud96}, 
neo-cortex \cite{Kisvarday93} and 
thalamus \cite{Bal93}.
It is important to make a detailed study on the
coupled HH neurons, which is the simplest but 
meaningful network unit.
The purpose of the present paper is to investigate
the spike-train responses of the coupled, excitable
HH neurons, which is treated as data-processing neurons
as in our previous study \cite{Hasegawa00}.
We consider the four kinds of recurrent couplings 
between a pair of HH neurons: the excitatory-excitatory (E-E),
inhibitory-inhibitory (I-I), excitatory-inhibitory (E-I), and
inhibitory-excitatory (I-E) couplings.
We apply, to the coupled HH neurons,
the two kinds of spike-train inputs:
one is the transient inputs consisting
of clustered $M$ impulses for the finite duration ($M$: integer)
and the other is the sequential spike train with the
constant ISI. The transient and asymptotic property of
their response will be investigated.

Our paper is organized as follows:
In the next Sec. II, we describe a simple neural system
consisting of neurons, axons, synapses and dendrites, which are 
adopted for our numerical calculation.
We present the calculated results in Sec. III:
the response of the coupled HH neurons to
the transient, clustered impulses is discussed in Sec. IIIA
and that to the sequential spike-train input in Sec. IIIB. 
The dependences of the 
distribution of the output ISIs on the coupling strength
and the time delay are studied.
The final Sec. IV is devoted to conclusion and discussion.

\begin{center}
{\bf II. ADOPTED MODEL}
\end{center}

We adopt a simple neural system
consisting of a pair of neurons which is numbered 1 and 2.
The neurons which are described by the HH model with
identical parameters,
are coupled with the time delay
of $\tau_{jk} \: (j,k=1,2)$ for an impulse 
propagating from the neuron $k$ 
to the neuron $j$.
This delay time is the sum of conduction times
through the axon and dendrite.
It has been reported that real biological synapses
exhibit temporal dynamics of depression or
potentiation during neuronal computation
\cite{Abbott97}\cite{Tsodyks98}.
We, however, treat the synapse as a static
unit for a simplification of our calculation.
The synapse with the coupling strength $C_{jk}$ 
is excitatory or inhibitory, and
it is assumed to be described by the alpha function [Eq.(7)].

Dynamics of the membrane potential $V_{j}$ of
the coupled HH neuron {\it j} (=1, 2) 
is described by the non-linear DDEs given by 

\begin{equation}
\bar{C} d V_{j}(t)/d t = -I_{j}^{\rm ion}(V_{j}, m_{j}, h_{j}, n_{j})  
+ I_{j}^{\rm ext} + I_{j}^{\rm int}(\{V_{k}(t-\tau_{jk})\}),
\end{equation}
where $\bar{C} = 1 \; \mu {\rm F/cm}^2$ is the capacity of the membrane.
The first term of Eq.(1) expresses the ion current given by
\begin{equation}
I_{j}^{\rm ion}(V_{j}, m_{j}, h_{j}, n_{j}) 
= g_{\rm Na} m_{j}^3 h_{j} (V_{j} - V_{\rm Na})
+ g_{\rm K} n_{j}^4 (V_{j} - V_{\rm K}) 
+ g_{\rm L} (V_{j} - V_{\rm L}).
\end{equation}
Here the maximum values of conductivities 
of Na and K channels and leakage are
$g_{\rm Na} = 120 \; {\rm mS/cm}^2$, 
$g_{\rm K} = 36 \; {\rm mS/cm}^2$ and
$g_{\rm L} = 0.3 \; {\rm mS/cm}^2$, respectively; 
the respective reversal potentials are   
$V_{\rm Na} = 50$ mV, $V_{\rm K} = -77$ mV and 
$V_{\rm L} = -54.5 $ mV.
The gating variables of Na and
K channels, $m_{j}, h_{j}$ and $n_{j}$,
are described by 
\begin{equation}
d m_{j}/d t = - (a_{mj} + b_{mj}) \: m_{j} + a_{mj},
\end{equation}
\begin{equation}
d h_{j}/d t = - (a_{hj} + b_{hj}) \: h_{j} + a_{hj},
\end{equation}
\begin{equation}
d n_{j}/d t = - (a_{nj} + b_{nj}) \: n_{j} + a_{nj}.
\end{equation}
The coefficients of 
$a_{mj}$ and $b_{mj}$ {\it etc.} are expressed in terms of 
$V_{j}$ (their explicit expressions 
having been given in Refs.\cite{Hasegawa00}\cite{Park96}) and
then the variables $V_{j}$, $m_{j}$, $h_{j}$ 
and $n_{j}$ are coupled. 

The second term in Eq.(1) denotes the external input currents given by
\begin{equation}
I_{j}^{\rm ext} = I_{{\rm s}j} + A_{s} \delta_{j1} \sum_{n} \alpha(t-t_{in}),
\end{equation}
with the alpha function $\alpha(t)$ given by
\begin{equation}
\alpha(t) = (t/\tau_{\rm s}) \; e^{-t/\tau_{\rm s}} \:  \Theta(t),
\end{equation}
The first term ($I_{{\rm s}j}$) in Eq.(6) is the dc current which determines
whether the neuron is excitable or periodically oscillating.
Its second term expresses the postsynaptic current
which is induced by
the  presynaptic spike-train input applied 
to the neuron 1, given by
\begin{equation}
U_{\rm i}(t) = V_{\rm a} \: \sum_n  \: \delta (t - t_{{\rm i}n}).
\end{equation}
In Eqs.(6)-(8),
$\Theta (t)=1$ for $x \geq 0$ and 0 for $x < 0$;
$A_{\rm s}=g_{\rm s} (V_{\rm a}-V_{\rm s})$,
$g_{\rm s}$ and $V_{\rm s}$ stand 
for the conductance and reversal potential,
respectively, of the synapse;
$\tau_{\rm s}$ is the time constant relevant 
to the synapse conduction, which is assumed to be
$\tau_{\rm s}=2$ msec; 
$t_{{\rm i}n}$ is the $n$-th firing time of 
the spike-train inputs defined recurrently by
\begin{equation}
t_{{\rm i}n+1} = t_{{\rm i}n} + T_{{\rm i}n}(t_{{\rm i}n}),
\end{equation}
where the input ISI $T_{{\rm i}n}$ is generally a function
of $t_{{\rm i}n}$.
For the constant input ISI of $T_{{\rm i}n}=T_{\rm i}$, 
$t_{in}$ is given by
$t_{{\rm i}n} = n T_{\rm i}$ 
for an arbitrary integer $n$.

When the membrane potential of the {\it j}-th neuron $V_{j}(t)$ oscillates,
it yields the spike-train output, which may be expressed by
\begin{equation}
U_{{\rm o}j}(t) = V_{\rm a} \: \sum_m \: \delta (t - t_{{\rm o}jm}),
\end{equation}
in a similar form to Eq.(8), 
$t_{{\rm o}jm}$ being the $m$-th firing time 
when $V_{j}(t)$ crosses $V_{z}$ = 0.0 mV from below.
The output ISI is given by
\begin{equation}
T_{{\rm o}jm} = t_{{\rm o}jm+1} - t_{{\rm o}jm}.
\end{equation}

The third term in Eq.(1) which 
expresses the interaction between the two neurons,
is assumed to be given by
\begin{equation}
I_{j}^{\rm int}(\{V_{k}(t-\tau_{jk})\})
=   \sum_{k (\neq j)} \sum_{m} \: C_{jk} 
\: \alpha(t-\tau_{jk}-t_{{\rm o}km}).
\end{equation}
The positive and negative $C_{jk}$ denote 
the excitatory and inhibitory couplings, respectively.
Depending on the signs of $C_{jk}$, we consider the
following four types of couplings (Fig.1):

(a) $C_{21} > 0$ and $C_{12} > 0  \:\:\:\:$  (E-E coupling),

(b) $C_{21} < 0$ and $C_{12} < 0  \:\:\:\:$  (I-I coupling),

(c) $C_{21} > 0$ and $C_{12} < 0  \:\:\:\:$  (E-I coupling),

(d) $C_{21} < 0$ and $C_{12} > 0  \:\:\:\:$  (I-E coupling).

\noindent
In order to reduce the number of parameters, we assume that 
$\mid C_{21} \mid =\mid C_{12} \mid=c \: A_{\rm s}$ and 
$\tau_{21}=\tau_{12}=\tau_d$.

As for the functional form of the coupling term of
$I_{j}^{\rm int}(\{V_{k}(t-\tau_{jk})\})$,
Foss, Longtin, Mensour and Milton \cite{Foss96}
adopt a simpler form given by 
\begin{equation}
I_{j}^{\rm int}(\{V_{k}(t-\tau_{jk})\})
=   \sum_{k (\neq j)}  \: \mu_{jk} \: V_{k}(t-\tau_{jk}),
\end{equation}
taking no account of the synapse, 
where $\mu_{jk}$ is the coefficient of the synaptic coupling.
They discuss the memory storage of the pattern in output spike trains,
injecting the input information by the initial function,
$V(t)$ for $t \in [-\tau_d, 0)$, whereas in our calculation
input information is given by $I^{\rm ext}_{j}$ [Eq.(6)].

Differential equations given by Eqs.(1)-(5) including
the external current and 
the coupling given by Eqs.(6)-(12) are solved 
by the forth-order Runge-Kutta method.
The calculation for each set of parameters is performed
for 2 sec (200,000 steps) otherwise noticed
by the integration time step of 0.01 msec
with double precision.
The initial conditions for the variables are given by
\begin{equation}
V_{j}(t)= -65 \:\: \mbox{\rm mV}, m_{j}(t)=0.0526,  
h_{j}(t)=0.600, 
n_{j}(t)=0.313, \:\: \mbox{\rm for} \:\: j=1, 2
\:\:  \mbox{at} \: t=0,
\end{equation}
which are the rest-state solution of a single
HH neuron ($c_{jk}=0$).
The initial function for $V_{j}(t)$,
whose setting is indispensable for
the delay-differential equation, is given by
\begin{equation}
V_{j}(t)= -65 \:\: \mbox{\rm mV} 
\:\: \mbox{\rm for} \:\: j=1,2
\:\:  \mbox{at} \: t \in [-\tau_d, 0).
\end{equation}
For an analysis of asymptotic solutions,
we discard results of initial 200 msec
(20,000 steps).

\vspace{0.5cm}
\begin{center}
{\bf III. CALCULATED RESULTS}
\end{center}

In the present study, we consider only the excitable HH neurons
by setting $I_{{\rm s}j}=0$ and $A_{\rm s}=40 \mu {\rm A/cm}^2$.
Our model has additional three parameters, $T_{\rm i}$, 
$\tau_{\rm d}$ and $c$. We treat them as free parameters to be changed 
for the four kinds of E-E, I-I, E-I, and I-E couplings
because the values of ISI and the time delay observed
in biological systems distribute 
in a fairly wide range \cite{note1}.

\begin{center}
{\bf A Transient Spike-Train Inputs}
\end{center}

Let us first investigate the response to the transient,
clustered spike-train
inputs consisting of $M$ impulses.
Figure 2(a) shows the example of
the time courses of input ($U_{\rm i}$),
output pulses ($U_{{\rm o}j}$),
the total postsynaptic current 
($I_{j}=I_{j}^{ext}+I_{j}^{int}$) and the membrane potential ($V_{j}$)
with $M=3$, $T_{\rm i}=20$, $\tau_{\rm d}=10$ msec and $c=1.0$
for the E-E coupling ($c > 0$).
The first external pulse
applied at $t=0$ yields the firing of the neuron 1
after the intrinsic delay of $\tau_{{\rm i}1} \sim 2$ msec. 
The emitted impulse propagates the axon and 
reaches the synapse of the neuron 2 after $\tau_{21}=10$ msec.
After a more delay of an intrinsic  $\tau_{{\rm i}2} \sim 2$ msec,
the neuron 2 makes the firing which 
yields the input current to the neuron 1 
after a delay of $\tau_{12}=10$ msec.
The input pulses trigger the continuous oscillation in
the coupled HH neurons with the output
ISI of $T_{{\rm o}}=24.1$ msec. 
Filled and open circles in Fig. 2(b) denote the time dependence 
of the output ISI of the neuron 1 and 2, respectively.
We note that $T_{{\rm o}1}$ and  $T_{{\rm o}2}$ start from 
the values of 20.00 and 19.96 msec, respectively, 
and soon become the fixed value of 24.10 msec.

Figure 3(a) shows 
the result for the I-I coupling 
with the parameters same as in Fig. 2(a). 
After $\tau_{21}=10$ msec of the firing of the neuron 1, the
inhibitory input arrives at the neuron 2,
which fires by the inhibitory rebound process
with an additional delay of $\tau_{{\rm i}1} \sim 15$ msec
beside a delay of $\tau_{12}=20$ msec of the axon.
Subsequently the neuron 1 induced by the
recurrent impulse coming from the neuron 2,
fires after an total time delay of 
$\tau_{21}+\tau_{{\rm i}2} \sim 25$ msec
due to delays of the axon and the inhibitory rebound.
Figure 3(b) shows that 
the output ISIs of the neuron 1 and 2
start from the values of 20.00 and 21.04 msec, respectively,
and asymptotically approach the oscillating two values of 
24.33 and 24.45 msec.

In the cases shown in Figs. 2 and 3, the delay of 
$\tau_{\rm d}=10$ msec is shorter than
the duration of input pulses (40 msec).
On the contrary, when the delay time becomes larger 
than the duration time of clustered inputs, 
the situation is changed.
Figures 4 shows the time courses of the membrane 
potentials of the neuron 1 with the four couplings  
for $M=3$, $T_{\rm i}=20$, $\tau_{\rm d}=50$ msec and $c=1.0$
(hereafter we consider only the output of the 
neuron 1).
We note that the triggered oscillation continues, 
keeping the original three-impulse form in all the couplings.
Then the coupled HH neurons play a role of memory storage
\cite{Foss96}\cite{Ikeda87}.

The period of the induced oscillation $T_{\rm p}$
shown in Fig. 4 is essentially determined by
the total delay time of the feedback loop $T_{\rm fb}$
given by
\begin{equation}
T_{\rm p} \sim T_{\rm fb} 
= 2 \tau_{\rm d} + \tau_{{\rm i}1} + \tau_{{\rm i}2}
\end{equation}
where $\tau_{{\rm i}j}$ denotes the intrinsic time delay of 
the neutron $j$.
It is noted that $\tau_{{\rm i}j}$ depends on whether
an input is excitatory or inhibitory:
$\tau_{{\rm i}j}$ is about 2-3 msec for the excitatory input while 
it is about 14-15 msec for the inhibitory input.
Our simulation shown in Fig. 4
yields $T^{\rm EE}_{\rm p}=105$, $T^{\rm II}_{\rm p}=129$,
$T^{\rm EI}_{\rm p}=T^{\rm IE}_{\rm p}=117$ msec for 
the E-E, I-I, E-I and I-E couplings, respectively,
which follows Eq.(16) with 
$\tau_{\rm i}$ = 2.5 and 14.5 msec for the excitatory and 
inhibitory synapses, respectively ($\tau_{\rm d}=50$ msec). 

\begin{center}
{\sl 1. The coupling-strength dependence}
\end{center}

Figure 5 shows the coupling-strength dependence of the period and
the output ISIs of the asymptotic solution
in the cases of the four couplings
for  $M=3$, $T_{\rm i}=20$ and $\tau_{\rm d}=50$ msec \cite{note2}.
The period of the oscillation triggered by input pulses has 
little dependence on $c$, as far as $c$ is
larger than the critical value, $c_{\rm cr}$, above which
the oscillation continues.
In the case of $M=3$, $T_{\rm i}=20$ and $\tau_{\rm d}=50$ msec, 
for example, we get $c_{\rm cr}=0.20$ for the E-E coupling and 
0.42 for the I-I, E-I and I-E couplings:
note that at $0.44 \leq c \leq 0.76$ in the latter cases, 
we get the two-impulse output
against the three-impulse input.
This is because the firing by the inhibitory rebound requires 
a considerable synaptic magnitude \cite{Hasegawa00}.

\begin{center}
{\sl 2. The time-delay dependence}
\end{center}

Next we study how the output ISIs are determined.
When the feedback time $T_{\rm fb}$ is larger than the duration
of clustered impulses
({\it i.e.} $T_{\rm fb}=2 \tau_{\rm d} + \tau_{{\rm i}1} + \tau_{{\rm i}2} 
> (M-1) \:T_{\rm i}  $), we get two values of
$T_{{\rm o}}$ given by
\begin{equation}
T^{(1)}_{\rm o} = T_{\rm i}, \:\:\: 
T^{(2)}_{\rm o} = T_{\rm fb} - (M-1) \:T_{\rm i}
= 2 \tau_{\rm d} + \tau_{{\rm i}1} + \tau_{{\rm i}2} 
- (M-1) \:T_{\rm i}.    
\end{equation}
On the other hand, when the feedback time is shorter than
input-pulse duration ($2 \tau_{\rm d} + \tau_{{\rm i}1} + \tau_{{\rm i}2} 
< (M-1) \:T_{\rm i}$),
we get
\begin{equation}
T^{(1)}_{\rm o} = T_{\rm i} \;\Theta(M-3), \:\:\: 
T^{(2)}_{\rm o} = \mid \ell \:T_{\rm fb} - m \:T_{\rm i} \mid
= \mid \ell\: (2 \tau_{\rm d} + \tau_{{\rm i}1} + \tau_{{\rm i}2}) 
- m \:T_{\rm i} \mid,  
\end{equation}
where integers $\ell$ and $m$ satisfy 
$1 \leq \ell \leq [(M-1)T_{\rm i}/T_{\rm fb}]+1$ and
$0 \leq m \leq M-1$, $[\:\cdot\:]$ is the Gauss sign 
and $T^{(1)}_{\rm o}$ is vanishing for $M \leq 2$.

We have obtained the $\tau_{\rm d}$ 
dependence of the output ISI from the asymptotic solution
in our simulations.
Results for clustered inputs of $M=3$
are shown in Fig. 6 (a)-6(d),
where filled circles 
denote the values of $T_{\rm o}$ for a given $\tau_{\rm d}$
\cite{note2}.
Figure 6(a) shows the result for inputs with 
$T_{\rm i}=20$ msec in the E-E couplings.
We try to analyze the calculated $T_{\rm o}$
with the use of Eqs.(17) and (18).
Three dashed lines in Fig. 6(a) are given by $2\tau_{\rm d}+5$,
$2\tau_{\rm d}-15$ and $2 \tau_{\rm d}-35$, which are
obtained by Eqs.(17) and (18) with 
$\tau_{{\rm i}1}=\tau_{{\rm i}2}=2.5$
and $T_{\rm i}=20$ msec.
Dashed lines in Fig. 6(b) are given by $2\tau_{\rm d}+29$,
$2\tau_{\rm d}+9$ and $2 \tau_{\rm d}-11$, which are
obtained by Eqs.(17) and (18) with 
$\tau_{{\rm i}1}=\tau_{{\rm i}2}=14.5$
and $T_{\rm i}=20$ msec.
Similar analysis is made for
the E-I and I-E couplings, whose results
are expressed by dashed lines in Figs. 6(c) and 6(d). 
Agreement of the dashed lines with filled circles
in Fig. 6(a) and 6(b) are fairly good.
On the contrary, the agreement between
them is not satisfactory in Fig. 6(c) and
6(d) for $T_{\rm i}=50$ msec.
The deviation of filled circles from the dashed lines is
due to the effects of non-linearly of the HH neurons.

The time courses of the membrane potentials for
$M=1-5$ are plotted in Fig. 7 for $T_{\rm i}=20$,
$\tau_{d}=50$ msec and $c=1.0$. In all the couplings, the 
form of input pulse is preserved and
the coupled neurons play a role of memory storage.

It is possible to control
the switching of the oscillations by input
pulses.
Figure 8 demonstrates such a switching in the E-I coupled
HH neurons.  The oscillation is triggered by the
external input $U_{1}$ at $t=0$.
When we apply the external excitatory input at $t=543$ msec,
it cancels out
the recurrent inhibitory input to the neuron 1 and suppress
the firing of the neuron 1. Then 
the oscillation is switched off by the control input pulse
at $t=543$ msec.
At $t=700$ mec, the oscillation is again triggered
by input pulse. 
In this example, the control pulse is applied
to the neuron 1.  We can alternatively switch off the oscillation
by applying the inhibitory (excitatory) input to
the neuron 2 in the E-I (I-E or I-I) coupling.
These show the feasibility of the on-off switching of 
the oscillation in the simplest neuronal unit.

\begin{center}
{\bf B. Sequential Spike-Train Inputs}
\end{center}

Next we discuss the response to the sequential spike-train.
Our calculations in I show that
when an isolated HH neuron ($c=0$) receives the sequential
inputs with the constant ISI of $T_{\rm i}$,
it behaves as a low-pass filter:
it emits the spike train with $T_{\rm o}>10$ msec
for $T_{\rm i}<12$ msec while for $T_{\rm i}>12$ msec
its output ISI is given by $T_{\rm o}=T_{\rm i}$
(see Fig.7 of Ref.\cite{Hasegawa00}).
This response may be modified when the coupling
is introduced to a pair of HH neurons.
Figures 9(a) and 9(b) show
the time courses of input ($U_{\rm i}$),
output ($U_{{\rm o}j}$),
the total postsynaptic current 
($I_{j}=I_{j}^{\rm ext}+I_{j}^{\rm int}$) and the 
membrane potential ($V_{j}$)
for $T_{\rm i}=20$ msec, $\tau_{\rm d}=10$ msec and $c=1.0$
for the E-E and I-I couplings, respectively.
They should be compared with Figs. 2(a) and 3(a)
for the transient spike-train inputs
with the same values of $T_{\rm i}$ and $\tau_{\rm d}$.
The output ISI in Fig. 2 is 24.1 msec while that
in Fig. 9(b) is 20 msec which is the entrained value
with input ISI.
Although $T_{\rm o}$ in Fig. 3 are two
values of 24.3 and 24.5 msec,
we get, in Fig. 9(b), {\it ten} values of 20.6, 21.2,
37.3, 19.2, 22.1, 20.6, 21.7, 35.7, 19.9 and 21.8 msec.

\begin{center}
{\sl 1. The coupling-strength dependence}
\end{center}

The response behavior of the coupled neurons strongly 
depends on the parameters of $c$, $\tau_{\rm d}$ and $T_{\rm i}$.
Figure 10(a)-10(d) show the examples of the $c$ dependence of the
distribution of $T_{\rm o}$ of the E-E, I-I, E-I and I-E
couplings for various parameters \cite{note2}.
We note that, as increasing the $c$ value,
the distribution of the output ISIs show the bi(multi)furcation,
as commonly observed in systems with the delayed feedback
\cite{Ikeda87}.
In order to investigate the phenomenon in more detail,
we show, in Fig. 11(a), the enlarged plot 
for the range of $0.6 \leq c \leq 1.2$
sandwitched by the dotted, vertical lines in Fig. 10(a).
Figure 11(b) is the enlarged plot of Fig. 10(d)
for the narrow range of $0.3 \leq c \leq 1.0$.
These figures  clearly show the bi(multi)furcation 
with many windows.

A cycle whose output ISIs almost
continuously distribute, is expected to be chaotic 
although in the strict sense,
the distribution of $T_{\rm o}$s never 
becomes continuous because they
are {\it quantized} by the integration time step of 0.01 msec.
Among many candidates of chaos-like behavior in Figs. 10 and 11, 
we pay our attention
to the result of $c=0.95$ in Fig. 11(a), for which
the Lorentz plot (return map) of its $T_{\rm o}$ is shown 
in Fig. 12(a)
(calculations are performed for 20 sec of two million steps).
The output ISIs seem to distribute on the folded ring.
When these points are connected by lines in the chronological
order, the inside of the ring is nearly filled by them.
In order to examine the property of this cycle,
we calculate the correlation dimension $\nu$ 
given by \cite{Grassberger83}
\begin{equation}
\nu= \lim_{\epsilon \rightarrow 0} 
\frac{\log \: C(\epsilon)}{\log \: \epsilon},
\end{equation}
with
\begin{equation}
C(\epsilon)=N^{-2} \sum_{m,n=1}^{N} 
\Theta(\epsilon - \mid \mbox{\boldmath $X_{m}$}
- \mbox{\boldmath $X_{n}$}   \mid),
\end{equation}

\begin{equation}
\mbox{\boldmath $X_{m}$}= (T_{{\rm o}m}, \: T_{{\rm o}m+1}, 
\:...., \: T_{{\rm o}m+k-1} ),
\end{equation}
where $C(\epsilon)$ is the correlation integral, {\boldmath $X_{m}$}
is the $k$-dimensional vector generated by $T_{{\rm o}m}$
$N$ the size of data, and $\Theta(\cdot)$ the Heaviside function.
Figure 12(b) shows the $\log C(\epsilon)$-$\log \epsilon$ plot
for various $k$ calculated for the cycle shown in Fig. 12(a)
with $N \sim 1200$.
We note that $C(\epsilon)$ behaves as 
$C(\epsilon) \propto \epsilon^{\nu}$ with the correlation
dimension of $\nu =0.94\pm0.02$
for small $\epsilon \:\: (0.01 = e^{-4.6} < \epsilon < e^{0})$.
The non-integral $\nu$ implies that this cycle may be chaos,
the related discussion being given in Sec. IV.

\begin{center}
{\sl 2. The time-delay dependence}
\end{center}

A simple generalization of the discussion 
given in the previous Sec. IIIA2 yields that the output ISIs
for the sequential spike-train input with the constant $T_{\rm i}$ 
are given by

\begin{equation}
T_{\rm o} = \mid \ell\: T_{\rm fb} - m \:T_{\rm i} \mid
= \mid \ell \: (2 \tau_{\rm d} + \tau_{{\rm i}1} + \tau_{{\rm i}2}) 
- m \:T_{\rm i} \mid,  
\end{equation}
where $\ell, m$ are integers.
In the above analysis, we do not take into account the absolute
refractory period and the merging of
recurrent inputs with external inputs at synapse, 
which eliminate some of the
candidates of $T_{\rm o}$ obtained from Eq.(22) for many choices
of a pair of $\ell$ and $m$, as will seen shortly.

Figure 13(a)-13(d) show the $\tau_{\rm d}$ dependence of the
distribution of $T_{\rm o}$ in the asymptotic solution
for the sequential inputs \cite{note2}.
We try to analyze the obtained distribution of $T_{\rm o}$
with the use of Eq.(22) by adopting 
$\tau_{{\rm i}j}$=2.5 and 14.5 msec for the excitatory 
and inhibitory couplings, respectively, as was made
in Sec. IIIA2.
Dashed curves in Fig. 13(c)-13(d) denote the lines generated
by Eq.(22) for these values of $\tau_{{\rm i}j}$
and $T_{\rm i}=50$ msec with a choice of a pair
of integers $(\ell, m)$ shown beside the lines.
Although many lines can be drawn for various choices of the integers,
we plot only results with $\ell=1$ in order to avoid
a disfigurement by them. 
Some of filled circles representing
the distribution of $T_{\rm o}$ in Figs. 13(c)-13(d)
may be explained by dashed lines. Most of filled circles, however,  
cannot be well understood by this analysis due to
the effects not included in our simple analysis.
Furthermore our analysis with the use of Eq.(22) does not
work at all for the results shown in Fig. 13(a) and 13(b),
which plot the distribution of $T_{\rm o}$ 
in the E-E and I-I couplings for inputs
with $T_{\rm i}=20$ msec.
We note that the result shown in Fig. 13(a) (13(b))
for the sequential inputs is quite
different from that shown in Fig. 6(a) (6(b)) for the
transient inputs although both the calculations
adopt the same parameters of $T_{\rm i}$ and $c$.

In order to see more the detailed structure
of the bi(multi)fucation,
we show, in Fig. 14(a), the enlarged plot  
for the range of $21 < \tau_{\rm d} < 26$ msec
between the dotted, vertical lines in Fig. 13(a).
Figure 14(b) is the enlarged plot of Fig. 13(c) for
the narrow range of $33 < \tau_{\rm d} < 38$ msec.
They clearly show the bi(multi)furcation as changing 
$\tau_{\rm d}$.

\vspace{0.5cm}
\begin{center}
{\bf VI. CONCLUSION AND DISCUSSION}
\end{center}

We have performed numerical investigation on the
spike-train responses of a pair of HH neurons
with four kinds of the recurrent E-E, I-I, E-I and I-E
couplings.
The response of the coupled, excitable HH neurons to
the transient, clustered impulses and to the sequential
inputs shows a  rich of variety.
The former input triggers
the synchronous oscillation 
in the coupled HH neurons, which may
play a role of memory storage. 
The response to the latter sequential input strongly 
depends
on the coupling strength and the time delay, yielding
bi(multi)furcation, multistabilty and chaotic behavior.

We have applied the two types of inputs of
the transient and  sequential spike-train impulses.
On the theoretical point of view, the latter
is taken as the limit of $M \rightarrow \infty$
of the former.
Figure 15(a) and 15(b) show the $T_{\rm i}$ dependence of
the distribution of $T_{\rm o}$
for the transient and sequential inputs, respectively,
with $\tau_{\rm d}=50$ msec and $c=1.0$ in the E-E coupling.
Dashed curves in Fig. 15(a) are given by the equations 
shown beside the lines which are obtained from Eqs.(17)
and (18), as mentioned in Sec. IIIA2.
On the other hand, dashed lines in Fig. 15(b) are obtained
with the use of Eq.(22) for a pair of integers $(\ell, m)$ shown 
in the brackets. 
Note that dashed lines given by $T_{\rm i}$,
$\pm T_{\rm i} \mp 105$ and $\pm 2T_{\rm i} \mp 105$ in Fig. 15(a)
are identical with those in Fig. 15(b) with 
$(\ell, m)$=(0,1), (1,1) and (1,2), respectively.
It is apparent that
the distribution of $T_{\rm o}$ in Fig. 15(a) is not
the same as that in Fig. 15(b), but they are partly similar.
For example, we obtain for $T_{\rm i}=20$ msec, 
$T_{\rm o}=19.6$, 20.0 and 64.54 msec
in Fig. 15(a) whereas only 20.0 msec in Fig. 15(b).
In order to understand this difference, we plot, in Fig. 16,
the time dependence of $T_{\rm o}$ for this set
of parameters by changing the $M$ value.
For $M=3$, $T_{\rm o}$ oscillates with the values of
19.6, 20.0 and 64.54 msec, 
as mentioned above.
The calculated $T_{\rm o}$ for $M$=4 are 18.7, 19.8, 20.0, and 45.6 msec, 
and those for $M=5$ are 19.9, 20.0 and 24.2 msec.
For $M=10$, $T_{\rm o}$ remains 20 msec until $t \sim 200$ msec, after which
$T_{\rm o}$ oscillates with the values of
19.9, 20.0 and 24.2 msec.
In the limit of $M \rightarrow \infty$ corresponding to the 
sequential inputs, the state with $T_{\rm o}=20$ msec
continues from $t=0$ to $\infty$.
Thus as increasing $M$, the time region of $T_{\rm o}=20$ msec
is increased.

Figure 17 shows the similar plot of the time
dependence of the distribution of $T_{\rm o}$
for various $M$
with $T_{\rm i}=20$, $\tau_{\rm d}=13.75$ and $c=0.95$,
for which the sequential input leads to the chaotic behavior,
as was discussed in Sec. IIIB1 (see Fig.12(a)).
In the case of $M=3$, we get the oscillation in $T_{\rm o}$
which asymptotically approaches the value of 15.97 msec.
In the case of $M=10$ (50), the chaotic behavior is
realized at $0 \leq t \siml < 180$
($0 \leq t \siml < 980$) msec during the application of inputs.
After inputs are switched off,
the output ISI gradually approach the fixed value of
15.97 msec.
In the limit of $M \rightarrow \infty$, the chaotic oscillation
eternally continues.

We have shown in Sec. IIIB1 that 
the cycle of the output ISIs shown
in Fig. 12(a) may be chaos because 
its correlation dimension of $\nu \sim 0.94$ is
derived from the $\log C(\epsilon)$-$\log \epsilon$ 
plot in Fig. 12(b).
This is not surprising because the response of
{\it single} HH neurons to some kinds of
external inputs may be chaotic 
\cite{Aihara84}\cite{Matsumoto84}\cite{Hasegawa00}.
In particular, it has been shown in I that the response of 
a single HH neuron may be chaos when  the ISI of
the spike-train input is modulated by
the sinusoidal signal \cite{Hasegawa00}:
\begin{equation}
T_{\rm i}(t) = d_0 + d_1 \: \sin(2 \pi t/T_p).
\end{equation}
Figure 18(a) shows the Lorentz plot of the output 
ISIs of the single HH neuron receiving 
sequential inputs with ISIs
sinusoidally modulated by Eq.(23) 
with $d_0=2 d_1=20$ and $T_p=100$ msec (see Fig.7(d) of I).
We note that $T_{\rm o}$s distribute on the deformed ring.
From the $\log C(\epsilon)$-$\log \epsilon$ plot (not shown)
of this cycle, we get its correlation dimension
of $\nu \sim 1.04$.
We apply this spike-train input to the E-E coupled
HH neurons with $\tau_{\rm d}=10$ msec and $c=1.0$, 
whose Lorentz plot is shown in Fig. 18(b).
Its structure is rather different from that shown 
in Fig. 18(a). 
Actually the correlation dimension of this cycle for
the coupled HH neurons
is $\nu \sim 1.83$, which is different from and larger than
$\nu \sim 1.04$ of the cycle shown in Fig. 18(a) 
for the single HH neuron.  
From similar calculations for the coupled HH nurons, 
we obtain the correlation dimensions of
$\nu \sim 0.95$ for $\tau_{\rm d}=5$ msec and $c=1.0$, and
$\nu \sim 1.03$ for $\tau_{\rm d}=10$ msec and $c=0.5$.
These results clearly show that
the correlation dimension of the output ISIs
depend not only 
on the model parameters ($c$ and $\tau_{\rm d}$)
of the coupled HH neurons but also 
on the correlation dimension
of input ISIs ($\nu_{\rm i}=0$ for the constant ISI and
$\nu_{\rm i}=1$ for the sinusoidally modulated ISI).
We expect that spike-train inputs with larger correlation dimension
lead to spike-train outputs with larger $\nu$.
One of the disadvantages of the present calculation
of the correlation dimension
is a lack of the data size of $N \sim 1200$
with million-step calculations.
A more accurate analysis  
requires a larger size of data and then
a computer with the larger 
memory storage.

Next we discuss the time correlation 
$\Gamma_{12}(\tau)$ between the 
membrane potentials, $V_1$ and $V_2$, of the 
neurons 1 an 2, defined by
\begin{equation}
\Gamma_{12}(\tau) = \int_{t_a}^{t_b} V_1(t) \: V_2(t+\tau) 
\: {\rm d}t,
\end{equation}
where $t_a=1000$ and $t_b=2000$ msec 
are adopted for our calculation.
Figure 19(a) shows the result
for the case of the sequential input
to the E-E coupled HH neurons with $T_{\rm i}=20$, 
$\tau_{\rm d}=10$ msec and $c=1.0$
(see Fig. 9(a) for the
time courses of $V_1$ and $V_2$).
In this case we obtain the constant $T_{\rm o}=20$ msec as 
was discussed in Sec. IIIB, and then
$\Gamma_{12}(\tau)$ shown in Fig. 19(a)
has peaks at $\tau=12.04+20 n$ msec ($n$: integer)
with the period of 20 msec, as expected.
We are interested in the time correlation
for the case when the distribution
of $T_{\rm o}$ is chaotic.
Results for such cases are shown in Figs. 19(b) and 19(c).
We have discussed in Sec. IIIB1 that the cycle of $T_{\rm o}$
depicted in Fig. 12 may be chaotic.
Figure 19(b) shows the result of this case
for the E-E coupling with $T_{\rm i}=20$, 
$\tau_{\rm d}=13.75$ msec and $c=0.95$.
We note that $\Gamma_{12}(\tau)$ has peaks
at $\tau$=0.0, 16.07, $\sim 3.2$, 48.17, 62.71,...
msec with the period of about 16 msec, which is
the sum of $\tau_{\rm d}$ and $\tau_{i1}$.
More evident peaks are found in Fig. 19(c) showing also
the chaotic case discussed in the preceding paragraph: 
the E-E coupled
neurons receiving the sinusoidal inputs given by Eq.(23) with 
$d_0=2 d_1=20$, $T_{\rm p}=100$, $T_{\rm i}=20$, 
$\tau_{\rm d}=10$ msec and $c=1.0$ [see Fig. 18(b)].
We note peaks in $\Gamma_{12}(\tau)$ at $\tau$=0.0, 12.72, 
25.30, 37.88, 50.43, ... msec
with the period of about 12.6 msec.
It is interesting that the synchronization is well
preserved between the coupled HH neurons even 
when the distribution of their output ISIs
shows the chaotic behavior.

A fairly large variability ($c_{\rm v}=0.5 \sim 1.0$) has
been reported for spike trains of non-bursting cortical
neurons in V1 and MT of monkey \cite{Softky92}.
It is possible that when the appreciable 
variability in neuronal signals
is taken into account in our calculations,
much of the fine structures in the $c-$ and 
$\tau_{\rm d}$-dependent distributions of $T_{o}$ will be washed out.
In order to study this speculation, we apply the spike-train input
with ISI whose distribution is given by the gamma distribution
defined by \cite{Hasegawa00}
\begin{equation}
P(T) = s^r \;\; T^{r - 1} \;\; e^{- s T}/\; \Gamma(r),
\end{equation}
where $\Gamma \;(r)$ is the gamma function.
The average of input ISI is given by
$\mu_{\rm i} =  r/s$, its root-mean-square (RMS)  by
$\sigma_{\rm i} = \sqrt{ r}/s$ and its variability by
$c_{\rm vi} = \sigma_{\rm i}/\mu_{\rm i}=1/\sqrt{r}$.
Figure 19 shows the $\tau_{\rm d}$ dependence of 
the mean ($\mu_{\rm o}$) and 
RMS values ($\sigma_{\rm o}$) of the
output ISIs for $c_{\rm vi}=0.0$ (dashed curves)
and $c_{\rm vi}=0.43$ (solid curves) 
with $T_{\rm i}=20$ msec and $c=1.0$.
Note that $\sigma_{\rm o}$ provides us with the measure
of the width of the distribution of $T_{\rm o}$.
The distribution for $c_{\rm vi}=0$ has a fine
structure reflecting the strong $\tau_{\rm d}$ 
dependence of $T_{\rm o}$ [see Fig. 13(a)]. 
This fine structure is, however, 
washed out for $c_{\rm vi}=0.43$, as expected.
Detailed calculations of the response of the coupled HH neurons
to stochastic 
spike-train inputs are now under way and will be published 
elsewhere.

\section*{Acknowledgements}
This work is partly supported by
a Grant-in-Aid for Scientific Research from the Japanese 
Ministry of Education, Science and Culture.



\begin{figure}
\caption{
Adopted pairs of HH neurons with 
(a) the E-E, (b) I-I, (c) E-I, and (d) I-E couplings, open 
and filled circles denoting excitatory and inhibitory synapses,
respectively.
}
\label{fig1}
\end{figure}

\begin{figure}
\caption{
The time dependence of 
(a) the transient input ($U_{\rm i}$), 
output ($U_{{\rm o}j}$),
the total postsynaptic current ($I_{j}$) and the
membrane potential ($V_{j}$), and (b) the output ISI ($T_{{\rm o}j}$)
with $M=3$, $T_{\rm i}=20$ and $\tau_{\rm d}=10$ msec
in the E-E couplings. 
The result of $V_2$ in (a) is shifted
downward by 200 mV and scales for $U_{\rm i}$, 
$U_{{\rm o}j}$ and $I_j$ 
are arbitrary.
}
\label{fig2}
\end{figure}

\begin{figure}
\caption{
The time dependence of 
(a) the transient input ($U_{\rm i}$), 
output ($U_{{\rm o}j}$),
the total postsynaptic current ($I_{j}$) and the
membrane potential ($V_{j}$), and (b) the output ISI ($T_{{\rm o}j}$)
with $M=3$ and $T_{\rm i}=20$ msec
in the I-I couplings. 
Same as in Fig. 2.
}
\label{fig3}
\end{figure}

\begin{figure}
\caption{
The time course of the input ($U_{\rm i}$) and membrane potentials ($V$)
for $M=3$, $T_{\rm i}=20$ and $\tau_{\rm d}=50$ msec in the 
four couplings, results of I-I, E-I and I-E couplings being shifted
downward by 200, 400 and 600 mV, respectively.
}
\label{fig4}
\end{figure}

\begin{figure}
\caption{
The coupling-strength dependence of 
the oscillation period $T_{\rm p}$
and the output ISIs $T_{{\rm o}}$ of the E-E (solid curve, open circles),
I-I (dashed curve, filled circles), and E-I and I-E couplings 
(dotted curve, open triangles) for the transient
inputs with $M=3$, $T_{\rm i}$=20 and $\tau_{\rm d}$=50 msec.
The left (right) ordinate is for $T_{\rm o}$ ($T_{\rm p}$).
}
\label{fig5}
\end{figure}

\begin{figure}
\caption{
The $\tau_{\rm d}$ dependence of the
distribution of $T_{{\rm o}}$  
of (a) E-E ($T_{\rm i}=20$ msec), (b) I-I ($T_{\rm i}=20$ msec), 
(c) E-E ($T_{\rm i}=50$ msec), and 
(d) I-E couplings ($T_{\rm i}$=50 msec)
for the clustered inputs with $M=3$
and $c=1$. The dashed lines  are expressed by the equations 
written beside the lines (see text).
}
\label{fig6}
\end{figure}

\begin{figure}
\caption{
The time course of the membrane potentials ($V_1$) for
$M=1-5$ for $T_{\rm i}=20$ and $\tau_{\rm d}$=50 msec 
with the E-E coupling.
}
\label{fig7}
\end{figure}

\begin{figure}
\caption{
The time course of the external input ($U_{1}$), the 
postsynaptic current ($I_{1}$), and the membrane potential
($V_{1}$) in the E-I coupling with $\tau_{\rm d}$=50 msec,
demonstrating the switching of the oscillation by $U_{1}$.
Scales for $U_{1}$ and $I_{1}$ are arbitrary.
}
\label{fig8}
\end{figure}

\begin{figure}
\caption{
The time course of input ($U_{\rm i}$), output ($U_{{\rm o}j}$),
the total postsynaptic current ($I_{j}$), and the
membrane potential ($V_{j}$) for sequential input 
with $T_{\rm i}=20$ and $\tau_{\rm d}=10$ msec
in the (a) E-E and (b) I-I couplings.
}
\label{fig9}
\end{figure}

\begin{figure}
\caption{
The distribution of $T_{{\rm o}}$ as a function of 
the coupling strength $c$
of (a) E-E ($T_{\rm i}$=20, $\tau_{\rm d}$=13.75 msec), 
(b) I-I ($T_{\rm i}$=20, $\tau_{\rm d}$=20 msec),
(c) E-I ($T_{\rm i}$=20, $\tau_{\rm d}$=15 msec), 
and (d) I-E couplings ($T_{\rm i}$=20, $\tau_{\rm d}$=20 msec)
for the sequential inputs. 
The enlarged plots of the regions between 
dotted, vertical lines in (a)
and (c) are shwon in Fig. 11(a) and (b), respectively.
}
\label{fig10}
\end{figure}

\begin{figure}
\caption{
The enlarged plots of the distribution of $T_{{\rm o}}$
for (a) the E-E coupling with $T_{\rm i}=20$ and 
$\tau_{\rm d}=13.75$ msec (see Fig. 10(a)),
and for (b) the I-E coupling with $T_{\rm i}=20$ and 
$\tau_{\rm d}=20$ msec (see Fig. 10(c)).
The arrow in (a) denotes the $c$ value for which the Lorentz plot
is shown in Figs. 12(a).
}
\label{fig11}
\end{figure}

\begin{figure}
\caption{
(a) The Lorenz plot of $T_{{\rm o}}$ for $c=0.95$
with $T_{\rm i}=20$ and $\tau_{\rm d}$=13.75 in the E-E coupling, 
the computation being performed for 20 sec (two million steps).
(b) The correlation integral $C(\epsilon)$ of the cylce
shown in (a) as a function of 
$\epsilon$ in the log-log plot for various dimensions $k$,
the dashed line denoting $C(\epsilon) \propto \epsilon^{\nu}$ 
with the correlation dimension of $\nu=0.94$ (see text).
}
\label{fig12}
\end{figure}

\begin{figure}
\caption{
The $\tau_{\rm d}$ dependence of the
distribution of $T_{{\rm o}}$ of (a) E-E ($T_{\rm i}=20$ msec), 
(b) I-I ($T_{\rm i}=20$ msec), (c) E-E ($T_{\rm i}=50$ msec), and 
(d) I-I couplings  ($T_{\rm i}=50$ msec) for the sequential input
with $c=1.0$. 
Dashed lines in (c) and (d) are given by Eq.(22) with a pair of
integers $(\ell, m)$ shown beside the line.
The enlarged plots of the regions between dotted, vertical lines in
(a) and (c) are shown in Fig. 14(a) and (b), respectively.
}
\label{fig13}
\end{figure}

\begin{figure}
\caption{
The enlarged plot of the distribution of $T_{{\rm o}}$
for (a) $T_{\rm i}=20$ msec (see Fig. 13(a)) and 
(b) $T_{\rm i}=50$ msec (see Fig. 13(c)).
}
\label{fig14}
\end{figure}

\begin{figure}
\caption{
The $T_{\rm i}$ dependence of the
distribution of $T_{\rm o}$ for (a) the transient ($M=3$)
and (b) sequential spike-train input with $\tau_{\rm d}=50$ msec
and $c=1.0$.
Dashed lines are given by Eqs.(17), (18) and (22) (see text).
}
\label{fig15}
\end{figure}

\begin{figure}
\caption{
The time dependence of $T_{{\rm o}}$ for the clustered impulse
inputs with $M=3$, 10, 50 and $\infty$ with $T_{\rm i}=20$, 
$\tau_{\rm d}=50$ msec and $c=1.0$ in the E-E coupling,
results of $M$=3, 10 and 50 being shifted upward by 30, 20 and 10 mV, 
respectively.
The arrows denote the time
below which the inputs are continuously applied.
}
\label{fig16}
\end{figure}

\begin{figure}
\caption{
The time dependence of $T_{{\rm o}}$ for the clustered impulse
inputs with $M=3$, 10, 50 and $\infty$ with $T_{\rm i}=20$, 
$\tau_{\rm d}=13.75$ msec and $c=0.95$ in the E-E coupling.
Same as in Fig.16.
}
\label{fig17}
\end{figure}

\begin{figure}
\caption{
The Lorenz plots of $T_{{\rm o}}$ of (a) the single HH neuron
and (b) the E-E coupled HH neurons 
($\tau_{\rm d}=10$ msec and $c=1.0$)
receiving spike-train inputs whose ISIs are  modulated
by sinusoidal signal given by Eq.(23)
with $d_0=2 d_1=20$ and $T_p=100$ msec
 (see text).
}
\label{fig18}
\end{figure}

\begin{figure}
\caption{
The time correlation $\Gamma_{12}(\tau)$ between the membrane 
potentials of the neurons 1 and 2 for
(a) the constant-ISI input with $T_{\rm i}=20$, 
$\tau_{\rm d}=10$ msec and $c=1.0$,
(b) that with $T_{\rm i}=20$. $\tau_{\rm d}=13.75$ msec 
and $c=0.95$,
and (c) the sinusoidal input given by Eq. (23) 
with $d_1=2 d_2=20$, $T_{\rm p}=100$,
$\tau_{\rm d}=10$ msec and $c=1.0$.
The results of (b) and (c) are shifted downward
by 1.0 and 2.0, respectively (see text).
}
\label{fig19}
\end{figure}

\begin{figure}
\caption{
The $\tau_{\rm d}$ dependence of the mean ($\mu_{\rm o}$) and
rms ($\sigma_{\rm o}$) of output ISIs of the E-E copuled HH neurons
($c=1$) recieving sequential inputs of $T_{\rm i}=20$ msec
with $c_{\rm vi}=0.0$ (dashed curves) and 
$c_{\rm vi}=0.43$ (solid curves) (see text).  
}
\label{fig20}
\end{figure}


\begin{thebibliography}{99}
\bibitem{note2}
The intial conditions given by Eq.(14) and the intitial function
given by Eq.(15) are set every time at the beginning of
the calculation when the values of a given set of
parameters 
of $c$, $\tau_{\rm d}$ and $T_{\rm i}$ are changed.
\end{thebibliography}
\end{document}